\documentstyle[twoside,fleqn]{article}

\def\noi{\noindent}

\makeatletter

\renewcommand{\section}{\@startsection{section}{1}{0pt}%
        {-3.5ex plus -1ex minus -.2ex}{2.3ex plus .2ex}%
        {\large\bf\protect\raggedright}}

\renewcommand{\subsection}{\@startsection{subsection}{2}{0pt}%
        {-3ex plus -1ex minus -.2ex}{1.4ex plus .2ex}%
        {\normalsize\bf\protect\raggedright}}

\renewcommand{\thesubsubsection}%
        {\arabic{section}.\arabic{subsection}.\arabic{subsubsection}.}
\renewcommand{\@oddhead}{\raisebox{0pt}[\headheight][0pt]{%
   \vbox{\hbox to\textwidth{\rightmark \hfil \rm \thepage \strut}\hrule}}}
\renewcommand{\@evenhead}{\raisebox{0pt}[\headheight][0pt]{%
   \vbox{\hbox to\textwidth{\thepage \hfil \leftmark \strut}\hrule}}}
\newcommand{\heads}[2]{\markboth{\protect\small\it #1}{\protect\small\it #2}}
\newcommand{\Acknow}[1]{\subsection*{Acknowledgement} #1}

\makeatother

\newcommand{\Title}[1]{\noindent {\Large #1} \\}
\newcommand{\Author}[2]{\noindent{\large\bf #1}\\[2ex]\noindent{\it #2}\\}
\newcommand{\Abstract}[1]{\vskip 2mm \begin{center}
     \parbox{16.4cm}{\small\noindent #1} \end{center}\bigskip}
\newcommand{\foom}[1]{\protect\footnotemark[#1]}
\newcommand{\foox}[2]{\footnotetext[#1]{#2}}
\newcommand{\email}[2]{\footnotetext[#1]{e-mail: #2}}

\newcommand{\sect}[1]{Sec.\,#1}

\def\nq{\hspace{-1em}}
\def\nqq{\hspace{-2em}}
\def\nhq{\hspace{-0.5em}}

\def\cm{\hspace{1cm}}

\newcommand{\sequ}[1]{\setcounter{equation}{#1}}

\def\beq{\begin{equation}}
\def\eeq{\end{equation}}
\def\bear{\begin{eqnarray}}
\def\al{&\nhq}
\def\lal{&&\nqq {}}               
\def\bearr{\begin{eqnarray} \lal}
\def\ear{\end{eqnarray}}
\def\earn{\nonumber \end{eqnarray}}

\def\tst{\textstyle}

\def\nn{\nonumber\\ {}}
\def\nnv{\nonumber\\[5pt] {}}
\def\nnn{\nonumber\\ \lal }
\def\nnnv{\nonumber\\[5pt] \lal }
\def\yy{\\[5pt]}

\def\eql{\al =\al}

\def\eqdef{\stackrel{\rm def}{=}}
\def\e{{\,\rm e}}
\def\d{\partial}

\def\sign{{\,\rm sign\,}}
\def\diag{{\,\rm diag\,}}
\def\dim{{\,\rm dim\,}}
\def\const{{\rm const}}

\def\half{{\tst\frac{1}{2}}}
\def\then{\ \Rightarrow\ }

\def\DAL{\raisebox{-1.6pt}{\large $\Box$}\,}
\newcommand{\vars}[1]{\left\{\begin{array}{ll}#1\end{array}\right.}

\def\intl{\int\limits}
\def\wider{\vphantom{\int}}

\renewcommand{\theequation}{\thesection\arabic{equation}}
\catcode`\@=11 \@addtoreset{equation}{section}\catcode`\@=12
\def\rank{\mathop{\rm rank}\nolimits}
\def\eps{\varepsilon}
\def\A{{\cal A}}
\def\K{{\cal K}}
\def\M{{\cal M}}
\def\RR{{\cal R}}
\def\S{{\cal S}}
\def\TH{T_{\rm H}}
\def\oG{{\overline G}}
\def\oI{{\overline I}}
\def\oc{{\overline c}}
\def\od{{\overline d}}
\def\Nsq{N_s^2}
\def\sumn{\sum_{i=1}^{n}}
\def\sums{\sum_s}
\def\umx{u_{\max}}
\def\m{{\rm m}}
\def\ss{{\scriptstyle s}}
\def\R{{\sf R\hspace*{-0.85ex}\rule{0.1ex}{1.5ex}\hspace*{0.85ex}}}
\def\tolim{\mathop{\longrightarrow}\limits}
\def\pr{\mathstrut'}

\heads{K.A. Bronnikov, V.D. Ivashchuk and V.N. Melnikov}
       {The Reissner-Nordstr\"om Problem for Intersecting
        Electric and Magnetic $p$-branes}

\begin{document}
\thispagestyle{empty}
\twocolumn[
\noi \unitlength=1mm
\begin{picture}(174,8)
   \put(31,8){\shortstack[c]
       {RUSSIAN GRAVITATIONAL SOCIETY                \\
       INSTITUTE OF METROLOGICAL SERVICE             \\
       CENTER OF GRAVITATION AND FUNDAMENTAL METROLOGY}       }
\end{picture}
\begin{flushright}
                                         RGS-VNIIMS-97/15 \\
                                         gr-qc/9710054 \\
	{\it Grav. and Cosmol.} {\bf 3}, No. 3 (11), 1997
\end{flushright}

\vspace*{1.5cm}

\Title{THE REISSNER-NORDSTR\"OM PROBLEM \yy
FOR INTERSECTING ELECTRIC AND MAGNETIC $P$-BRANES}

\Author{K.A. Bronnikov\foom 1, V.D. Ivashchuk and V.N. Melnikov\foom 2}
{Centre for Gravitation and Fundamental Metrology, VNIIMS,\\
          3--1 M. Ulyanovoy St., Moscow 117313, Russia}

{\it Received 18 August 1997}

\Abstract
	{A multidimensional field model with (at most) one Einstein space of
	non-zero curvature and $n$ Ricci-flat internal spaces is considered.
	The action contains arbitrary numbers of dilatonic scalar fields
	$\varphi^a$ and antisymmetric forms $F_s$ of both electric and
	magnetic types, as they appear in the weak field limit of theories
	like M-theory (associated with $p$-branes).  The problem setting
	covers various models with field dependence on a single space-time
	coordinate, in particular, homogeneous cosmologies, static,
	spherically symmetric and Euclidean models. Exact solutions
	are obtained when the $p$-brane dimensions and the dilatonic couplings
	obey orthogonality conditions in the minisuperspace. For spherically
	symmetric solutions, conditions for black hole and wormhole existence
	are formulated. As in 4 dimensions, wormholes can only exist with
	negative energy density (such as pure imaginary scalar fields).
	For black holes, an analogue of no-hair theorems for $F$-forms is
	obtained; it is shown that even in spaces with multiple times
	a black hole may only exist with its unique, one-dimensional time;
	an infinite value of the Hawking temperature is explicitly shown
	to imply a curvature singularity at an assumed horizon, and such cases
	among extreme black-hole solutions are indicated.}

]    
\foox 1 {E-mail: kb@goga.mainet.msk.su. Present address: Departamento
   de F\'{\i}sica, Universidade Federal do Esp\'{\i}rito Santo, Vit\'oria
   --- CEP 29060--900, ES, Brazil, e-mail: kb@cce.ufes.br}
\email 2 {melnikov@fund.phys.msu.su}

\section{Introduction}

The properties of $p$-branes interacting with gravity have been recently
discussed by many authors. Such problems naturally emerge in bosonic
sectors of supergravitational models \cite{CJS,SS} and may be of interest in
the context of superstring and M-theories [3--7].

We continue our previous studies on this trend [8--14]
and consider multidimensional gravitational models containing several
coupled dilatonic scalar fields and antysymmetric forms \cite{IMO},
in the case of essential dependence of all fields on a single variable.
The model, outlined in \sect 2, describes generalized intersecting
$p$-branes (for different aspects of $p$-branes see \cite{St,D,AIR} and
references therein) in isotropic and anisotropic cosmological models,
static space-times with spherical and other symmetries, and the
corresponding Euclidean models.  In \sect 3, using the harmonic coordinate
condition and a $\sigma$-model representation \cite{IMO}, we reduce the
equations of motion to a Toda-like Lagrange system. In the simplest case of
orthogonal vectors in the exponents of the Toda potential we obtain exact
solutions.  \sect 4 briefly discusses the general properties of the
solutions in the cosmological and spherically symmetric cases, in
particular, the possible existence of wormhole configurations. In \sect 5,
among all spherically symmetric solution we select the subclass of
black-hole ones and give a general expression for their Hawking
temperature. The latter proves to be infinite in the extreme limit of
some of the holes, indicating that, in this limit, they become naked
singularities. In \sect 6 we formulate some restrictions for $p$-brane
configurations and, in particular, black holes in space-times with multiple
time coordinates.  As a by-product, we obtain a certain generalization of
the no-hair theorems (namely, that black holes are incompatible with the
so-called quasiscalar forms). Finally, the Appendix gives, under the
general conditions of the model of \sect 2, an explicit proof of the
existence of a naked singularity at a surface with infinite gravity;
applied to horizons, this means that an infinite Hawking temperature
indicates a singularity.

Special cases of the present models were recently studied
by a number of authors (\cite{LPX,AIV} and others); in particular,
spherically symmetric and cosmological models with a conformally invariant
generalization of the Maxwell field to higher dimensions were discussed
in Refs.\,[18--20];
in \cite{Br97} some integrable cases of
cosmological models with perfect fluid sources were indicated. We here
deal with the general class of coupled electric and magnetic
$p$-brane and dilatonic fields, but with no other material sources.

     For convenience we list some indices used in the paper and
     their corresponding objects:
\begin{description}
\item[]
     $M,\ N,... \mapsto $ coordinates of the $D$-dimensional Riemannian space
	$\M$;
\item[]
     $I,\ J,... \mapsto $ subsets of the set $I_0 = \{0,1,\ldots,n\}$;
\item[]
	$\e,\m \   \mapsto $ electric and magnetic type forms, respectively;
\item[]
	$s,s'      \mapsto $ $\e I$ or $\m I$, unified indices;
\item[]
	$a,\ b,... \mapsto $ scalar fields;
\item[]
	$i,\ j,... \mapsto $ subspaces of $\M$;
\item[]
	$m_i,\ n_i \mapsto $ coordinates in $\M_i$;
\item[]
	$A,\ B,... \mapsto $ minisuperspace coordinates.
\end{description}
	As usual, summing over repeated indices is assumed when one of them is
	at a lower position and another at an upper one.


\section{Initial field model}

	We start (like \cite{IMO}) from the action
\bearr                                                        \label{2.1}
     S = \frac{1}{2\kappa^{2}}
	             \intl_{\M} d^{D}z \sqrt{|g|} \biggl\{ {R}[g]  \nnn
	- \delta_{ab} g^{MN} \d_{M} \varphi^a \d_{N} \varphi^b
                                                    - \sum_{s\in \S}
 	\frac{\eta_s}{n_s!} \e^{2 \lambda_{sa} \varphi^a} F_s^2
                  \biggr\},
\ear
     in a $D$-dimensional (pseudo-)Riemannian manifold $\M$
	with the metric tensor $g_{MN}$; $|g| = |\det (g_{MN})|$;
	$\varphi^a$ are dilatonic scalar fields;
\beq
    \nhq	F_s =  dA_s = 	\frac{1}{n_s!} F_{s,\ M_1 \ldots M_{n_s}}
	           dz^{M_1} \wedge \ldots \wedge dz^{M_{n_s}}  \label{2.2}
\eeq
	are $n_s$-forms ($n_s \geq 2$) in $\M$;
	$\lambda_{sa}$ are coupling constants;
	$\eta_s = \pm 1$ (to be specified later); furthermore,
\beq                                                          \label{2.3}
	F_s^2 =  F_{s,\ M_1 \ldots M_{n_s}} F_s^{M_1 \ldots M_{n_s}},
\eeq
	$s \in \S$,  $a\in \A$,
	where $\S$ and $\A$ are non-empty finite sets.

	The Lagrangian (\ref{2.1}) gives the equations of motion
\bearr                                                        \label{2.4}
	G_{MN}\equiv R_{MN} - \frac{1}{2} g_{MN} R  =  T_{MN},
                   \\ \lal                                    \label{2.5}
   \DAL \varphi^a =
	\sum_{s \in \S} \eta_s \frac{\lambda_{sa}}{n_s!}
	\e^{2 \lambda_{sb} \varphi^b} F_s^2,  \\ \lal          \label{2.6}
          \nabla_{M}\left(\e^{2 \lambda_{sa} \varphi^a}
				F_s^{MM_2 \ldots M_{n_s}}\right)  =  0,
\ear
	where $\DAL$ and $\nabla$ are the Laplace-Beltrami and covariant
     derivative operators corresponding to $g$ and
\bearr                                                         \label{2.7}
	T_{MN} =  \sum_{a\in {\cal A}} T_{MN}[\varphi^a] +
          \sum_{s \in \S} \eta_s
          \e^{2 \lambda_{sa} \varphi^a} T_{MN}[F_s], \\ \lal
                                                               \label{2.8}
	T_{MN}[\varphi^a] =
              \d_{M} \varphi^a \d_{N} \varphi^a -
	     \frac{1}{2} g_{MN} \d_{P} \varphi^a \d^{P} \varphi^a, \\ \lal
    T_{MN}[F^s]                                               \nnn
    = \frac{1}{n_{s}!}  \biggl[ - \frac{1}{2} g_{MN} F_s^{2}
      + n_{s} F_{s,M M_2 \ldots M_{n_s}}
              F_{s,N}^{M_2 \ldots M_{n_s}}\biggr] .    \nnn  \label{2.9}
\ear

     We take the manifold $\M$ and the metric on it in the form
\bear                                                        \label{2.10}
	\M \eql  \R  \times \M_{0} \times \ldots \times \M_{n},\\
	ds^2 \eql w \e^{2{\alpha}(u)} du^2 +                    \label{2.11}
		        \sum_{i=0}^{n} \e^{2\beta_i(u)} ds_i^2,
\ear
     where $w=\pm 1$, $u$ is a selected coordinate, to serve as an argument
     for the unknown functions;
     $ ds_i^2  = g^i_{m_{i} n_{i}}(y_i) dy_i^{m_{i}}dy_i^{n_{i}} $
     are metrics on $\M_i$, satisfying the equation
\beq                                                         \label{2.12}
	R_{m_{i}n_{i}}[g^i ] = \xi_{i} g^i_{m_{i}n_{i}},
\eeq
	$m_{i},n_{i}=1,\ldots, d_{i}$ ($d_{i} = \dim M_i$);  $\xi_i= \const$.

	We assume each manifold $\M_i$, $i = 0,\ldots,n$,
        to be oriented and connected, so that one can correctly define
        their volume $d_i$-forms $\tau_i$ and signature parameters $\eps_i$:
\bear                                                        \label{2.13}
	\tau_i \eql \sqrt{|g^i(y_i)|}
             		\ dy_i^{1} \wedge \ldots \wedge dy_i^{d_i}, \nn
	\eps_i \eql \sign \det (g^i_{m_{i}n_{i}}) = \pm 1.
\ear

	For the dilatonic scalar fields we put
	$\varphi^a = \varphi^a(u)$, $a \in {\cal A}$.
	Let us now specify the set $\S$ in such a way as to include both
	electric and magnetic type $p$-branes with essential dependence of the
	field variables on $u$ only. For each non-empty subset
	$I = \{ i_1, \ldots, i_k \}$ ($i_1 < \ldots < i_k$)
     of the set of indices $I_0 = \{ 0, \ldots, n \}$ one can define an
	electric type $F$-form (\ref{2.2}), choosing the potential
	$A = A_{\e I}$ as
\beq
	A_{\e I} = \Phi_{\e I}(u)\tau_I, \cm                    \label{2.15}
	\tau_I \eqdef \tau_{i_1}\wedge \ldots \wedge \tau_{i_k}.
\eeq
	Thus the rank of the form $A_{\e I}$ coincides with
	$d(I) = \sum_{i\in I} d_i$, the dimension of the oriented manifold
	$\M_I = \M_{i_1}  \times  \ldots \times \M_{i_k}$. With (\ref{2.15}),
\beq
	F_{\e I} = dA_{\e I} = \dot\Phi_{\e I} du \wedge \tau_I,  \label{2.17}
\eeq
	where a dot denotes $d/du$.
	By construction and according to (\ref{2.17}),
\beq                                                       \label{2.18}
	n_{\e I} = \rank F_{\e I} = d(I) + 1,
\eeq
	so that the ranks of the forms $F_{\e I}$ are fixed by
	the manifold decomposition.

	In the conventional $p$-brane problem setting [5--7]
     it is supposed
	that one of the coordinates of $\M_I$ is time and the form (\ref{2.17})
	corresponds to a $(d(I)-1)$-brane  ``living'' in the remaining
	subspace of $\M_I$.

	A magnetic-type $F$-form may be defined as a form dual to some
	electric-type one, namely,
\beq                                                       \label{2.20}
	F_{\m I} = \e^{-2\lambda_{\m I a}\varphi^a}
	                          * [d\Phi_{\m I}\wedge\tau_I],
\eeq
	where $\Phi_{\m I} = \Phi_{\m I} (u)$ is a potential function,
     $\lambda_{\m I a}$ are coupling constants and $*$ is the Hodge operator:
\bearr
	(*F)_{M_1\ldots M_k} \eqdef \frac{\sqrt{|g|}}{k!}       \label{2.21}
	  \eps_{M_1\ldots M_kN_1\ldots N_{D-k}} F^{N_1\ldots N_{D-k}},\nnn
\ear
	$\eps_{\ldots}$ being the Levi-Civita symbol. Thus
\beq
	n_{\m I}=\rank F_{\m I} = D- \rank F_{\e I} = d(\oI),  \label{2.22}
\eeq
	$\oI \eqdef I_0 \setminus I$.
	Nonzero components of $F_{\m I}$
	correspond to coordinates of the subspaces $\M_i,\ i\in \oI$.

	Now, the set of indices $\S = \{s\}$ in (\ref{2.1}) will unify the
	electric and magnetic $F$-forms, the latter being also
	distinguished by the symbol $\chi_s = \pm 1$:
\bearr                                                      \label{2.23}
     \S = \{\e I\} \cup \{\m I\},        \qquad
	\chi_s = \vars  {
	                 +1 &{\rm for}\ F_{\e I},   \\
				  -1 &{\rm for}\ F_{\m I}.    }\nnn
\ear
	The number of elements in $\S$ is $|\S| = 2(2^{n+1}-1)$.

\medskip\noi
	{\bf Remark 1.}
	It is certainly possible to associate several, instead
	of one, electric and/or magnetic forms with each $I$; such a
	generalization is, however, straightforward and, making notations more
	cumbersome, changes actually nothing in the solution process.

\medskip\noi
	{\bf Remark 2.}
	If two or more forms $F_s$ have equal ranks $r$, they may be treated
	as different components of the same $r$-form, provided the
	coupling constants coincide as well. Such a situation is
	sometimes called ``composite $p$-branes" \cite{AIR}. There is just one
	exception: one must avoid unification of forms having only one
	noncoinciding index (which may happen if there is more than one
	1-dimensional subspaces among $M_i$), since in this case there emerge
	nonzero off-block-diagonal energy-momentum tensor (EMT) components,
	while the Einstein tensor in the l.h.s. of (\ref{2.4}) is
	block-diagonal.  Though, two or several such off-diagonal terms may in
	principle compensate each other, thus avoiding the prohibition. See
	more details in Ref.\,\cite{IM5}.
\medskip

	The problem setting as described embraces various classes of models
	where field variables depend on a single coordinate, such as
\begin{description}
\item[(A)]
	cosmological models, both isotropic and anisotro\-pic, where the $u$
	coordinate is timelike, $w=-1$, and some of the factor
	spaces (one in the isotropic case) are identified with the physical
	space;
\item[(B)]
	static models with various spatial symmetries (spherical, planar,
	pseudospherical, cylindrical, toroidal) where $u$ is a spatial
	coordinate, $w=+1$, and time is selected among $\M_i$;
\item[(C)]
	Euclidean models with similar symmetries, or models with a Euclidean
	``external" space-time; $w=+1$.
\end{description}

	A simple analysis shows that a positive energy density
	$-T^t_t$ of the fields $F^I$ is achieved in all Lorentzian models
	with the signature $(-++\ldots +)$
	if one chooses in (\ref{2.1}) $\eta_s = 1$ for all $s$, as is done
	in most works dealing with $F$-forms. In more general models, with
	arbitrary $\eps_i$, the requirement $-T^t_t >0$ is fulfilled if
\beq                          \wider
     \eta_{\e I} = - \eps(I)\eps_t(I),\cm
      \eta_{\m I}= - \eps(\oI) \eps_t(\oI),        	\label{2.25}
\eeq
	where $\eps(I)\eqdef \prod_{i\in I}\eps_i$; the quantity $\eps_t(I)=1$
	if the time ($t$) axis belongs to the factor space $\M_I$ and
     $\eps_t(I) = -1$ otherwise. Thus if $\eps_t(I) =1$, we are dealing
     with a genuine electric or magnetic field (see (\ref{2.17}) and
     (\ref{2.20})), while otherwise the $F$-form behaves as an effective
	scalar or pseudoscalar in the physical subspace. The latter happens, in
	particular, in isotropic cosmologies where nonzero spatial vectors
	would violate the isotropy. Such forms will be called quasiscalar.
	For example: if $z^M$, $M=0,1,2,3$ describe the external
	static space-time, $z^0=t$, $z^1=u$ and $M>3$ label extra dimensions,
	then, say, a form with nonzero $F_{015}$ is true electric,
	$F_{235}$ is true magnetic, $F_{156}$ is electric quasiscalar and
	$F_{0235}$ is magnetic quasiscalar.

	In the solution process we preserve, for generality,
	arbitrary sign factors, including $\eta_s$.

	We also use the following notations for the logarithms of
	volume factors of the subspaces of $\M$:
\bearr
	\sum_{i=0}^{n} d_i \beta_i \equiv \sigma_0, \cm
	\sum_{i=1}^{n} d_i \beta_i \equiv \sigma_1, \nnn \cm
     \sum_{i\in I}  d_i \beta_i \equiv \sigma(I).   \label{2.26}
\ear


\section{Solutions}

	Let us, as in \cite{Br73}, choose the harmonic $u$ coordinate
	($\DAL u = 0$), such that
\beq                                                         \label{3.1}
	\alpha (u)=\sigma_0 (u).
\eeq
	The Maxwell-like equations (\ref{2.6})  are easily solved and give
	(with (\ref{3.1})):
\bear                                                  \label{3.2}
	F_{\e I}^{uM_1\ldots M_{d(I)}}
		\eql \frac{Q_{\e I}}{\sqrt{|g_I|}}
		     \e^{-2\alpha - 2\vec \lambda_{\e I}\vec\varphi}
                      \eps^{M_1...M_{d(I)}},                          \\
	F_{\m I,\, M_1\ldots M_{d(\oI)}}                  \label{3.3}
		\eql Q_{\m I} \eps_{M_1...M_{d(\oI)}} \sqrt{|g_\oI|},
\ear
	where $|g_I| = \prod_{i\in I} |g^i (y_i)|$,
	the constants $Q_s$ are charges and arrows denote summing in $a$.
	These solutions lead to the following form of the EMTs (\ref{2.9}):
\def\Qie{Q_{\e I}}
\def\Qim{Q_{\m I}}
\bear                                                      \label{3.4}
   \e^{2\alpha}T_M^N (F_{\e I})
	  \eql \frac{w}{2}\eps(I) \eta_{\e I}\Qie^2
	      \e^{2\sigma(I) -4\vec \lambda_{\e I}\vec \varphi}  \nnn
          \quad \times \diag\bigl(+1,\ [+1]_I,\ [-1]_\oI \bigr);   \nn
   e^{2\alpha}T_M^N (F_{\m I})
	  \eql \frac{1}{2}\eps(\oI) \eta_{\m I}\Qim^2 \e^{2\sigma(I)}  \nnn
	\quad \times \diag\bigl(-1,\ [-1]_I,\ [+1]_\oI \bigr),
\ear
	where the first place on the diagonal belongs to $u$ and the symbol
	$[f]_J$ means that the quantity $f$ is repeated along the diagonal for
     all indices referring to $\M_i,\ i\in J$. On the other hand, the scalar
	field EMTs (\ref{2.8}) are
\beq                                                         \label{3.5}
	\e^{2\alpha}T_M^N (\varphi^a)
	    = \frac{w}{2} \bigl(\dot\varphi{}^a)^2 \diag(+1,\ [-1]_{I_0}\bigr).
\eeq

	To solve the remaining field equations,
	let us adopt the following assumptions:
\begin{description}
\item[(i)]
	Among all $\M_i$, only $\M_0$, with $d_0>1$,
	may have a nonzero curvature, while others
	are Ricci-flat, i.e., $\xi_i=0$ for $i >0$ and (properly normalizing
	the scale factor $\beta_0$)
\beq
	\xi_0 = (d_0-1)K_0, \cm  K_0 = 0,\ \pm 1.              \label{3.6}
\eeq
\item[(ii)]
	Neither of $I$ such that $\Qie\ne 0$ or $\Qim\ne 0$ contains the index
	0, that is, neither of the $p$-branes under consideration involves the
	subspace $\M_0$.
\end{description}

	Let us denote by $\S_*$ the set of $s$ with nonzero charges, i.e.,
	$\S_* = \{s\mid Q_s \ne 0\}$. In what follows, all sums and products in
	$s$ will be performed over the set $S_*$ without special indication.

	One can notice that each constituent of the total EMT (\ref{2.7})
	possesses the property
\beq
	T_u^u + T_z^z =0                                         \label{3.7}
\eeq
	if $z$ belongs to some $\M_i,\ i\in\oI$. Due to Assumptions (i) and
	(ii), a coordinate $z$ from $M_0$ is such a coordinate for all
	$s\in \S_*$ and hence the relation (\ref{3.7}) holds for the total EMT.
	As a result, the corresponding combination of the Einstein equations
	(\ref{2.4}) can be integrated:
\bearr
	G_u^u + G_z^z  \nnn
       = -w\e^{-2\alpha}                     \label{3.8}
	 \bigl[\ddot \alpha -\ddot \beta_0
	       - wK_0(d_0 -1)\e^{2\alpha - 2\beta_0} \bigr] = 0, \nnnv
	\e^{\beta_0 - \alpha} = (d_0-1) S(wK_0,\ k,\ u),
\ear
	where $k$ is an integration constant (IC) and we have introduced the
	notation
\bear
	S(1,\ h,\ t) \eql \vars{ h^{-1} \sinh ht, \quad & h>0,\\
				                        t,       & h=0,\\
				     h^{-1} \sin ht,     & h<0; }    \nn
	S(-1,\ h,\ t)\eql h^{-1} \cosh ht; \cm  h> 0;            \nn
	S(0,\ h,\ t) \eql \e^{ht}, \cm\qquad    h\in \R.         \label{3.9}
\ear
	Another IC is suppressed by properly choosing the origin of the $u$
	coordinate.

	With (\ref{3.8}) the $D$-dimensional line element may be written in the
	form
\bear
	ds^2 \eql \frac{\e^{-2\sigma_1/\od}}{[\od S(wK_0,k,u)]^{2/\od}} \nnn
	\nq \times
     \biggl[ \frac{w\, du^2}{[\od S(wK_0,k,u)]^2} + ds_0^2\biggr]
				+ \sumn \e^{2\beta_i}ds_i^2     \label{3.10}
\ear
	where $\od \eqdef d_0-1$.

	It is now helpful to represent the remaining field equations
	using the so-called $\sigma$-model
	(minisuperspace) approach (its more general form is described in
	\cite{IM4}).

	Let us treat the whole set of unknowns ${\beta_i(u),\,
        \varphi^a (u)}$ as a real-valued vector function $x^A (u)$ (so that
	$\{A\} = \{1,\ldots,n\} \cup \A$) in an $(n+|\A|)$-dimensional
        vector space $V$.  One can then verify that the field equations for
	$\beta_i$ and $\varphi^a$ coincide with the equations of motion
	corresponding to the Lagrangian of a Euclidean Toda-like system
\bear                                                        \label{3.12}
	L       \eql \oG_{AB}\dot x^A\dot x^B-V_Q (y),   \nnv
	V_Q (y) \eql  \sums \chi_s \theta_s Q_s^2 \e^{2y_s}
\ear
	where the nondegenerate symmetric matrix
\bearr                                                  \label{3.13}
    (\oG _{AB}) = \pmatrix {
  	      G_{ij}&       0      \cr
	      0     &  \delta_{ab} \cr }, \qquad
		G_{ij} = \frac{d_id_j}{d_0-1} + d_i \delta_{ij} \nnn
\ear
	defines a positive-definite metric in $V$; the sign factors $\theta_s$
	are
\beq
	\theta_{\e I} = \eta_{\e I} \eps(I); \cm
	\theta_{\m I} = w \eta_{\m I} \eps(\oI)                   \label{3.14}
\eeq
	and the functions $y_s(u)$ are defined in the following way:
\beq                                                            \label{3.15}
	y_s =
	(y_{\e I}, y_{\m I}) = \sigma(I_s) -\chi_s \vec\lambda_s\vec \varphi.
\eeq

	The Lagrange equations due to (\ref{3.12}) must be supplemented
	by an ``energy" integral that follows from the ${u \choose u}$
	component of the Einstein equations (\ref{2.4}):
\bear                                                         \label{3.16}
	E \eql \frac{1}{\od}\dot\sigma_1^2 + \sumn d_i \dot\beta_i^2
	          + \delta_{ab}\dot\varphi^a \dot\varphi^b + V_Q (y)    \nn
	     \eql  G_{AB}\dot x^A \dot x^B + V_Q (y)=\frac{d_0}{d_0-1}K, \\
        K \eql \vars {
                      k^2 \sign k, & wK_0 = 1; \\
                      k^2,         & wK_0 = 0,\ -1.       }
\earn
	where the IC $k$ has appeared in (\ref{3.8}).

	The functions (\ref{3.15}) can be represented as scalar products in $V$:
\bearr                                                          \label{3.17}
	y_s (u)   = Y_{s,A}  x^A,    \qquad
	(Y_{s,A}) = (d_i\delta_{iI}, \ \  -\chi_s \lambda_{sa}),\nnn
\ear
     where
	$\delta_{iI} \eqdef \sum_{j\in I}^{}\delta_{ij}$
	is an indicator of $i$ belonging to $I$ (1 if $i\in I$ and 0 otherwise).

	The contravariant components of $Y_s$ are found using the matrix
	$\oG^{AB}$ inverse to $\oG_{AB}$:
\bear                                                       \label{3.18}
	(\oG ^{AB})\eql \pmatrix{
		G^{ij}&      0      \cr
	0     &\delta^{ab}  \cr }, \qquad
	G^{ij}=\frac{\delta^{ij}}{d_i}-\frac{1}{D-2}, \nnn \\ \label{3.19}
	(Y_s{}^A) \eql
	\biggl(\delta_{iI}-\frac{d(I)}{D-2}, \ \ -\chi_s \lambda_{sa}\biggr)
\ear
	and the scalar products of different $Y_s$, whose values are of primary
	importance for the integrability of our system, are
\bearr
	Y_{s,A}Y_{s'}{}^A = d(I_s \cap I_{s'})                 \label{3.20}
					- \frac{d(I_s)d(I_{s'})}{D-2}
			+ \chi_s\chi_{s'} \vec \lambda_s \vec \lambda_{s'}.\nnn
\ear

	Let us now adopt two further assumptions making it possible to entirely
	integrate the field equations:
\begin{description}
\item[(iii)]
	$n + |\A| \geq |\S_*|$, i.e., the number of functions $y_s$ does not
	exceed the number of equations.
\item[(iv)]
	The vectors $Y_s$, $s \in \S_*$, are mutually orthogonal with
	respect to the metric $\oG_{AB}$, that is,
\bear                                                       \label{3.21}
\nq	Y_{s,A}Y_{s'}{}^A \eql \delta_{ss'}\big/ \Nsq, \nn
\nq	1\big/ \Nsq \eql
	  d(I)\biggl[1- \frac{d(I)}{D-2} \biggr] + {\vec \lambda_s}^2 >0.
\ear
\end{description}

	Under these assumptions the functions $y_s(u)$
	obey the decoupled equations
\beq                                                         \label{3.22}
	\ddot y_s = -\frac{\chi_s \theta_s Q_s^2}{N^2_s} \e^{2y_s},
\eeq
	whence
\beq                                                         \label{3.23}
	\e^{-y_s(u)} = \frac{|Q_s|}{N_s} S(-\chi_s\theta_s,\ h_s,\ u-u_s),
\eeq
	where $h_s$ and $u_s$ are ICs and the function $S(.,.,.)$ is defined in
	(\ref{3.9}). For the sought functions $x^A (u)$ we then obtain:
\beq
	x^A(u) = \sums \Nsq Y_s{}^A y_s(u) + c^A u + \oc^A,    \label{3.24}
\eeq
	where the vectors of ICs $c^A$ and $\oc^A$ satisfy the orthogonality
	relations $c^A Y_{s,A} = \oc^A Y_{s,A} = 0$, or
\bear
	c^i d_i\delta_{iI_s} - c^a\chi_s\lambda_{sa}   \eql 0, \nn
	\oc^id_i\delta_{iI_s} -\oc^a\chi_s\lambda_{sa} \eql 0.   \label{3.25}
\ear
	Specifically, the logarithms of scale factors $\beta_i$ and the scalar
	fields $\varphi^a$ are
\bear                                                       \label{3.26}
	\beta_i (u) \eql
                     \sums \biggl[
     \delta_{iI_\ss} - \frac{d(I_s)}{D-2}\biggr] \Nsq\, y_s (u)
							+c^i u + \oc^i, \nnn \\
								 \label{3.27}
	\varphi^a(u)\eql - \sums \lambda_{sa} N_s^2\, y_s(u) + c^a u + \oc^a,
\ear
	and the function $\sigma_1$ appearing in the metric (\ref{3.10}) is
\bearr                                                       \label{3.28}
	\sigma_1 = - \od\sums
	\frac{d(I_s)}{D-2}N_s^2\, y_s(u) + u\sumn c^i + \sumn \oc^i. \nnn
\ear

	Finally, the ``conserved energy" $E$ from (\ref{3.16}) is
\bearr
    \nhq	E = \sums \Nsq h_s^2\sign h_s + \oG_{AB}c^A c^B  \label{3.29}
	                = \frac{d_0}{d_0-1} k^2 \sign k.   \nnn
\ear

	The relations (\ref{3.1})--(\ref{3.3}), (\ref{3.8}), (\ref{3.10}),
	(\ref{3.23})--(\ref{3.29}), along with the definitions (\ref{3.9}) and
	(\ref{3.21}) entirely determine our solution, which is general under
	the above assumptions and involves all cases mentioned in items A, B, C
	at the end of \sect 2.

\medskip\noi
	{\bf Remark 3.} \ The above solutions evidently do not exhaust all
	integrable cases. Thus, some of the functions $y_s$ may coincide,
	reducing the number of the required orthogonality conditions
	and giving more freedom in choosing the set of input
	constants $d_i$ and $\lambda_{sa}$. Such a coincidence of different
	$y$'s is a constraint on the unknowns and the emerging
	consistency relations should lead to a reduction of the number of
	integration constants (e.g. coincidence of charges). In
	other words, it becomes possible to obtain less general solutions but
	for a more general set of input parameters. An example of such a
	situation is given in Ref.\,\cite{Pd97} discussing spherically
	symmetric solutions with intersecting electric and magnetic $p$-branes
	in the case of a single $F$-form. Other solutions must be yielded by
     the known methods of treating nontrivial Toda systems, see e.g.
	\cite{AA,IMnew}.

\medskip\noi
	{\bf Remark 4.} \ Electric and magnetic forms $F_{\e I}$ and
	$F_{\m I}$, specified on the same set $I$, in general, participate
	separately in the solution process, except for the case
	$\lambda_{\e Ia}= -\lambda_{\m Ia}$.
	In this case they form a common function $y_s$ and their charges appear
	in all relations (in Lorentzian models, under the conditions
	(\ref{2.25}) ) only in the combination $Q^2_{\e I} + Q^2_{\m I}$.
	Examples are evident, beginning with conventional Maxwell theory.


\section{Cosmology and spherical symmetry. Wormholes}

	The positive energy requirement (\ref{2.25}), fixing the input signs
	$\eta_s$, can be written as follows using the notations (\ref{3.14}):
\beq
	\chi_s\theta_s = -\eps_t(I).                      \label{4.1}
\eeq
	This condition essentially restricts the possible solution behaviour in
	particular cases. Thus, in cosmological models, when $u$ is a time
	coordinate, $\eps_t (I) = -1$, and $\M_0$ is naturally identified with
	the physical 3-space ($d_0=3$), (\ref{3.23}) gives all $y_s$ in terms
	of hyperbolic cosines, in full similarity to our previous work
	\cite{Greb}. Thus all conclusions, obtained there for the purely
	electric case, remain valid in the more general case of mixed electric
	and magnetic $p$-branes. In particular,
     all hyperbolic models with a 3-dimensional external space possess an
     asymptotic with a linear dependence of the external scale factor
	on the cosmic time $t$, while all internal scale factors
	and all scalar fields tend to finite limits.

	In static, spherically symmetric models, where $u$ is a radial
	coordinate, $w=+1$, $\M_0=S^{d_0}$, $K_0 = +1$, among other $\M_i$ there
	should be a one-dimensional subspace, say, $\M_1$, which may be
	identified with time, i.e. $\eps_1 = +1$. One can see that the sign
	factor $wK_0$ in (\ref{3.8}) is $+1$, while $-\chi_s\theta_s$ in
	(\ref{3.22}) is, due to (\ref{4.1}), $+1$ for normal electric and
	magnetic forms $F_I$ and is $-1$ for quasiscalar ones. Therefore
	(see (\ref{3.9})) the general solution combines
	hyperbolic, trigonometric and power functions, depending on the signs
	of the ICs $k$ and $h_s$, and a considerable diversity of behaviours is
	possible. One can reveal, however, that a generic solution
	possesses a naked singularity at the configuration centre, where
	$r(u) = \e^{\beta_0} \to 0$. Indeed, without loss of generality, the
	range of $u$ is $0 < u < \umx$, where $u=0$ corresponds to flat spatial
	infinity, while $\umx$ is finite when at least one of the constants $h_s$
	is negative and is infinite otherwise (by (\ref{3.29}), $k<0$ is only
	possible if some $h_s < 0$). In this case, $\umx$ is the smallest
	zero in the set of functions
\[
	\e^{-y_s} \sim \sin [|h_s| (u-u_s)]
\]
	Then it is clear from (\ref{3.26}) that all $\beta_i \to \pm\infty$,
	$i=1,\ldots,n$ as $u \to \umx$; on the other hand, according to
	(\ref{3.28}) $\sigma_1 \to \infty$, and the coordinate radius
\[
     r= \e^{\beta_0} = \bigl[\e^{-\sigma_1}\big/ S(+1,k,u)
	                                        \bigr]^{1/\od} \to 0
\]
	provided the denominator is finite. Hence the limit $u\to\umx$
	is the centre. Such singularities are similar to the
	Reissner-Nordstr\"om repulsive centre, with $g_{tt}\to \infty$
	and diverging energy of the respective electric or magnetic field
	component.

	Possible coincidences of zeros for different $\e^{-y_s}$ do not
	essentially alter the situation.

	Another generic case is that of $\umx = \infty$, when all $h_s \geq 0$.
	Then, as $u\to \infty$, the factors $\e^{\beta_i}$ behave like
	$\e^{k_i u}$, with constants $k_i$ of either sign and, moreover,
	possibly different signs for different $i$. Therefore again in the
	generic case we deal with a naked singularity, but this time it is not
	necessarily at the centre.

	An opportunity of interest is that of a (traversable,
	Lorentzian) wormhole, which realizes when $k<0$ and, in addition, the
	zero of $S(+1,k,u) \sim \sin |k|u$ (i.e. $u = \pi/|k|$) is smaller than
	those of $\sin [|h_s|(u-u_s)]$ for all $h_s$ which are negative. In
	this case, at $\umx = \pi/|k|$ we come through another spatial
	infinity, $\e^{\beta_0}\to \infty$, with finite limits of all other
	$\beta_i$ and $\varphi^a$. This opportunity, if realized, is also
	generic.

	A necessary condition for having a wormhole
	is $|k| > |h_s|$ for all $h_s$ which are negative. Suppose that they
	are all negative, which only favours the wormhole existence. The
	relation (\ref{3.29}) then leads to other {\sl necessary conditions in
	terms of the input parameters\/} of the model:
\bearr                                                    \label{4.2}
        \sums \Nsq > \frac{d_0}{d_0-1}          \nnn
          \then \ \sums \biggl[ d(I_s)
	\biggl(1-\frac{d(I_s)}{D-2}\biggr)\biggr]^{-1} > \frac{d_0}{d_0-1}.
\ear
	(In both inequalities the summing is performed over all $s$ able to give
     $h_s<0$, i.e., over only true electric and magnetic forms in the case of
     spherical symmetry. The second inequality is obtained from the
     first one by excluding $\lambda_{sa}$.)

	Thus wormhole solutions are, in principle, possible within this class
	of models. However, for systems satisfying the
	positive energy requirement (\ref{4.1}), with or without the scalar
	fields $\varphi^a$, one can prove that the existence of
	a wormhole throat is at variance with the properties of the total EMT
	in the 4-dimensional setting of the problem, just as in purely
	4-dimensional theory \cite{HV}; see more details in \cite{wh-new}. On
	the other hand, if one admits that at least one of $\varphi^a$ is pure
	imaginary rather than real, then the corresponding constants $c^a$
	become pure imaginary, and wormhole solutions are readily obtained
	(compare to \cite{Br73,b,i}).  Indeed, the term with $c^A$ in
	(\ref{3.29}) then loses its positive-definiteness, thus opening an
	opportunity to have arbitrarily large $k$ for given $h_s$; moreover,
	the coupling constants $\lambda_{sa}$ become pure imaginary as well,
	which enlarges the coefficients $N_s^2$ according to (\ref{3.21}) and
	also provides greater $|k|$ for given $h_s$.

	Cancelling the positive energy requirement for quasiscalar $F$-forms
	turns their $h_s>0$ into quantities of either sign, which also favours
	the construction of wormhole solutions. Consider, for example, the
	widely discussed 11-dimensional supergravity model (without scalar
	fields), namely, its solution with seven 2-branes ($d_s=3$,
	$d(I_s\cap I_{s'}) =1$, $s\ne s'$) \cite{Greb}. Among the seven
	$F$-forms only three can be electric or magnetic (since only three
	sets $I_s$ can simultaneously contain the time direction, in other
	words, only three branes can evolve with the time $t$).  This is
	insufficient for satisfying (\ref{4.2}); however, if the other four
	forms (they can be only quasiscalar) bear negative energy, wormhole
	solutions do exist.

	One can conclude that, in the present class of models, just as in
	4-dimensional theory, wormhole solutions appear only at the expense of
	explicitly violating the conventional energy requirements.


\section{Black holes}

	The brief analysis of the previous section has shown that solutions
	with $\umx < \infty$ describe either wormholes, or configurations with
	naked singularities. Let us now suppose that all $h_s \geq 0$ and try
	to select black-hole solutions.

	We first note that $h_s=0$ for some $s$ leads to singularities due to
	an uncompensated power behaviour of some $\e^{\beta_i}$.
	(We leave aside the special case of all $h_s=0$: then by (\ref{3.29})
	all $c^A=0$ and there appear numerous restrictions upon the input
	parameters if one tries to require regularity.)
	So we put $h_s > 0$, when all asymptotics are exponential.

	In search for black-hole solutions we make one further Ansatz, which is
	quite common in $p$-brane studies \cite{CT,AIV,Oh}:
\begin{description}
\item[(v)]
	The subspace $\M_1$ is time: $d_1=1$, $\eps_1=-1$ and $\delta_{1I} =1$
	for all $s = (I, \chi_s)$. (The latter means that all $p$-branes under
	consideration evolve with time.)
\end{description}

	Let us now require that all $|\beta_i| < \infty$, $i= 2,\ldots,n$
	(regularity of extra dimensions), $|\varphi^a| < \infty$ (regularity of
	scalar fields) and $|\beta_0| < \infty$ (finiteness of the spherical
	radius) as $u \to \infty$. This leads to the following constraints on
	the ICs:
\bear                                                    \label{5.1}
	c^A \eql \sums \Nsq Y_s{}^A h_s, \cm   A\ne 1;   \\   \label{5.2}
	c^1 \eql -k +  \sums \Nsq Y_s{}^1 h_s,
\ear
	where $A=1$ corresponds to $i=1$. Then, applying the orthonormality
	relations (\ref{3.25})  for $c^A$, we immediately obtain:
\bear
	h_s \eql k, \cm \forall\ s \in \S_*;
	                                               \label{5.3}  \yy
	c^A \eql k \sums \Nsq Y_s{}^A - k \delta^A_1.  \label{5.4}
\ear
	Surprisingly, the ``energy constraint" (\ref{3.29}) then holds
	automatically.

	It is now easy to verify that under these restrictions the solutions
	indeed describe black holes with a horizon at $u=\infty$. In
	particular, $g_{tt}\to 0$ as $u\to\infty$ and the light travel time
	$t = \int \e^{\alpha-\gamma} du $ diverges as $u\to\infty$. This family
	exhausts all black-hole solutions under the assumptions made, except
	maybe the limiting case $k=0$.

	For a clearer description and for comparison with other works let us
	apply the following transformation $u \mapsto R$:
\bearr
	\e^{-2ku} = 1 - \frac{2\mu}{R^{\od}},                    \label{5.5}
	         \qquad  \od \eqdef d_0-1, \qquad \mu \eqdef k\od.   \nnn
\ear
	Without loss of generality we adopt the boundary condition at spatial
	infinity: $\varphi^a \to 0$, $\beta_i \to 0$ as $u\to 0$.
	Then after the transformation (\ref{5.5}) the metric and the scalar
	fields acquire the form
\bear                                                         \label{5.6}
     ds^2 \eql \biggl(\prod_s H_s^{2\Nsq d(I_s)/(D-2)}\biggr)\nnn
        \times
	   \Biggl[ -\biggl(\prod_s H_s^{-2\Nsq}\biggr) (1-2\mu/R^\od)\ dt^2
                                                                     \nnnv
	   \cm + \biggl(\frac{dR^2}{1-2\mu/R^\od}
	                       + R^2 d\Omega^2_{\od+1}\biggr)           \nnn
     \cm + \sum_{i=2}^{n}\biggl(\prod_s H_s^{-2\Nsq \delta_{iI_s}}\biggr)
	                                ds_i^2 \Biggr],\yy       \label{5.7}
	 \varphi^a \eql \sums \Nsq \chi_s \lambda_{sa} \ln H_s,
\ear
	where $d\Omega^2_{\od+1}$ is the line element on a
	$(d_0=\od+1)$-dimensional sphere and the function $H_s$ (harmonic
	functions in $\R \times S^{\od+1}$) are
\bearr                                                        \label{5.8}
	H_s (R) = 1 + \frac{p_s}{R^\od}, \qquad
		   p_s \eqdef \sqrt{\mu^2 + \od^2 Q_s^2/\Nsq} - \mu  \nnn
\ear
	($N_s$ are defined in (\ref{3.21})).

	The active gravitational mass $M_{\rm g}$ is found from the
	asymptotic of $g_{tt}$ as $R \to \infty$:
\beq                                                         \label{5.9}
	G_N M_{\rm g} = \mu + \sums \Nsq \biggl[1-\frac{d(I_s)}{D-2}\biggr]p_s
\eeq
	where $G_N$ is Newton's gravitational constant.

	The extreme case when the mass is minimum for given charges $Q_s$,
	corresponds to the limit $\mu \to 0$ ($k \to 0$).
	In papers using the extreme limit as a basis for obtaining non-extreme
	solutions the parameter $\mu$ is sometimes called a deformation
     parameter \cite{CT,AIV}; we, on the contrary, obtain the extreme case
     as a limit of non-extreme black-hole solutions.
	It should be noted that in the extreme case the limit $R \to 0$ can
	be a singularity rather than an event horizon. Such cases can be
	revealed by infinite Hawking temperature, as shown in the Appendix.

	For the metric (\ref{5.6}) the Hawking temperature
	corresponding to $R^\od \to 2\mu$, calculated by standard methods
	(see (\ref{THa})), is
\beq                                                         \label{5.10}
     \TH = \frac{\od}{4\pi k_{\rm B} (2\mu)^{1/\od} }
		          \prod_s \biggl(\frac{2\mu}{2\mu + p_s}\biggr)^{\Nsq}.
\eeq
	Hence $\TH$ is zero in the limit $\mu\to 0$ if the parameter
	$\xi \eqdef \sums N_s^2 - 1/\od >0$, finite if $\xi=0$ and
	infinite if $\xi <0$. (A direct calculation of $\TH$ for the extreme
	configuration confirms these results.) In the latter case the horizon
	turns into a singularity in the extreme limit.

     On the other hand, the behaviour of $\TH$ as
	$\mu\to 0$ characterizes the black hole evaporation dynamics.
	Since $\TH$ depends on $d_0 = \od+1$ (which is 2 in the most physically
	plausible case) and on the norms $\Nsq$ which in turn depend on the
	$p$-brane properties, these features are potentially observable via the
	Hawking effect.

	A good example of the dependence of $\TH$ on the $p$-brane
	configuration is the (already mentioned at the end of \sect 4)
	11-dimensional supergravity, where it is possible to have at most
	three electro-magnetic 2-branes compatible with black holes. The norms
	are $N_s^2 = 1/2$ for each brane. Then, in the conventional ($S^2$)
	spherical case, $\od = d_0-1 = 1$ and it is clear from (\ref{5.10})
	that, in the extreme limit, $\TH$ tends to zero if there are three
	2-branes, to a finite limit if there are two and to infinity if there
	is only one brane. Another example is the dimension dependence of
	$\TH$ in the case of the conformally invariant generalized Maxwell
	field \cite{BrFa}.

\section{On models with multiple times}

	Some recent unification models assume that there can exist more than
     one time coordinate (see \cite{2times,im2t} and references therein).
	The above solutions involve any such cases since the signatures
	$\eps_i$ are there arbitrary.

	As has been mentioned in \sect 2, the positive energy requirement
	specifies the signs $\eta_s$ for $F$-forms ``living" in different
	subspaces of $\M$. This leads to a selection rule applied to
	``composite" $F$-forms, in addition to the restriction of Remark 2.
	Indeed, if an $F$-form, with a certain fixed factor $\eta$ in the
	action (\ref{2.1}), has several nonzero components attached to
	different $I$, then, in order to have positive energy of all such
	components, each of them taken separately should satisfy (\ref{2.25})
	with the same fixed $\eta$. For instance, all genuine electric-type
	$F_I$ must reside in $I_s$ with the same $\eps(I)$, which is the
	opposite of $\eps(I)$ for quasiscalar electric-type components, etc.

	Furthermore, if there is another time direction, it is natural to
	assume that some of the ``branes" evolve with this other time. Let us
	try to study this opportunity, cancelling Ansatz (v), i.e. allowing
	$\delta_{1I}=0$ for some $s$, and try to find a black-hole solution.
	A consideration quite similar to \sect 5 leads to $h_s=
	k \delta_{1I_s}$, so that $h_s=0$ for certain $s$. This in turn leads
	to power behaviour of some $y_s$, which is incompatible with regularity
	at $u\to\infty$.

     This reasoning has only used the fact that some $\delta_{1I} =0$,
     without reference to factor space signatures. Therefore we have here,
     as a by-product, {\sl an analogue of no-hair theorems}: there is no
     black-hole solution (in the family under consideration) with a nonzero
	quasiscalar $F$ field component. A similar theorem was proved in
	\cite{br95} for a simpler model ($D$-dimensional dilaton
	gravity), but for a general case when exact solutions were not
	obtained.

	Another conclusion is that ``black branes" ($p$-branes forming a
	black-hole configuration) must all evolve with the time $t$
	if the horizon is characterized by $g_{tt}=0$. Let us find out, whether
	it is possible to have a black hole with at least two times on equal
	footing, such that for another time $t'$ the corresponding space-time
	section be characterized by $g_{t't'}=0$. To this end, let us repeat the
	consideration of \sect 5 with $d_1>1$, but again with
	$\delta_{1I_s}=1$.  In a similar way one obtains:
\beq
	h_s=k,\cm
	c^A = \sums \Nsq Y_s{}^A - k \delta^A_1/d_1.             \label{6.1}
\eeq
	Substituting these expressions into the ``energy" constraint
        (\ref{3.29}), one
	obtains in the left-hand side, instead of $k^2 d_0/(d_0-1)$,
	the quantity
\beq
	k^2 \biggl( \frac{d_0}{d_0-1} + \frac{1}{d_1} -1\biggr).
\eeq
	Thus (\ref{3.29}) holds only for $d_1=1$.

	We conclude that even in a space-time with multiple time coordinates a
	black hole can only exist with its unique preferred (physical) time,
	while other times are not distinguished from extra spatial coordinates.

\section*{Appendix}

\renewcommand{\theequation}{A.\arabic{equation}}
\sequ{0}

\subsection*{The Riemann tensor and the Kretschmann scalar}

     The Riemann tensor $R^{MN}{}_{PQ}$ for the metric (\ref{2.11}) has
	the following nonzero components (see also \cite{IM-7}):
\bear                                                           \label{A1}
     R^{um_i}{}_{un_i} \eql -\delta^{m_i}_{n_i} \RR_i, \nn
	\RR_i\al\eqdef\al w\e^{-\beta_i-\alpha}
	              \bigl(\e^{\beta_i - \alpha}
	                            \dot\beta_i\bigr)\dot{\mathstrut}; \nn
     R^{m_i n_i}{}_{p_i q_i} \eql
           \e^{-2\beta_i}R^{m_i n_i}{}_{p_i q_i}[g^i]
			 - w \e^{-2\alpha}\dot {\beta}^2_i
						\delta^{m_i n_i}_{p_i q_i}, \nn
	\delta^{mn}_{pq} \al\eqdef\al
		       \delta^m_p \delta^n_q - \delta^n_p \delta^m_q;    \nn
	R^{m_i n_j}{}_{p_i q_j} \eql
			  -\delta^{m_i}_{p_i} \delta^{n_j}_{q_j} \RR_{ij}, \nn
				 \RR_{ij} \al\eqdef\al
	   w\e^{-2\alpha} \dot{\beta}_i \dot{\beta}_j, \cm i\ne j,
\ear
	and those obtained from (\ref{A1}) by evident index permutations.
	If $g^i$ is a metric of a constant curvature space, then
\beq
     R^{m_i n_i}{}_{p_i q_i}[g^i] = K_i\,
		\delta^{m_i n_i}_{p_i q_i}, \cm K_i = \const,    \label{A2}
\eeq
	and the third line of (\ref{A1}) may be rewritten as
\bear
	R^{m_i n_i}{}_{p_i q_i} \eql                             \label{A3}
	           \delta^{m_i n_i}_{p_i q_i}\, \tilde{\RR}_i, \nn
	\tilde{\RR}_i \al\eqdef\al \e^{-2\beta_i} K_i
			                     - w \e^{-2\alpha}\dot {\beta}^2_i.
\ear

	The Kretschmann scalar	$\K = R^{MN}{}_{PQ} R_{MN}{}^{PQ}$,
     whose infinite value at certain points of the space-time indicates
     a singularity, is
\beq                                                            \label{A4}
     \nq \K = \sum_{i=0}^{n}\Bigl [
	               4d_i\RR_i^2 + 2d_i (d_i-1) \tilde{\RR}^2_i \Bigr]
	   + 4 \sum_{i\ne j} d_i d_j \RR_{ij}^2.
\eeq

\subsection*{The Hawking temperature and singularities}

	One can prove the following sufficient (though not necessary)
	condition for a surface $u=\const$ in $\M$ to be a singularity:

\medskip\noi
	{\bf Statement A1.} \
     {\sl In space-times with the metric (\ref{2.11}),
	where $ds_i^2$ are metrics of constant curvature spaces,
	a surface $u=u^* = \const$ is a
	curvature singularity if, as $u\to u^*$, for some coefficient
	$\e^{\beta_i}\eqdef \e^{\gamma}$ it holds
\beq
	\e^{\gamma} < \infty, \cm                             \label{A5}
	          \e^{\gamma-\alpha}|\dot\gamma| \to \infty
\eeq
	(a dot is $d/du$), while in some neighbourhood of $u=u^*$
        the function $\gamma (u)$ is smooth and monotonic.}

\medskip\noi {\bf Proof.} \
	The structure (\ref{A4}) of the Kretschmann scalar indicates that,
	for $\K\to\infty$, it is sufficient that any single $\RR_i\to\infty$.
	To prove that it is indeed the case under the above conditions,
	let us use the fact that the quantities $\gamma$,
	$\e^{\gamma-\alpha}\dot\gamma$ and $\RR_i$ are insensitive to
	reparametrizations of the $u$ coordinate.
	Take a new coordinate $v$ such that $\alpha + \gamma \equiv 0$, then
\beq
	\e^{\alpha-\gamma}\gamma' = \half (\e^{2\gamma})\pr
	                           \tolim_{u\to u^*} \infty     \label{A6}
\eeq
	(a prime is $d/dv$) and $\RR_i = \half (\e^{2\gamma})\pr\pr$.
	Denote $\e^{2\gamma}= f(v)$, $(\e^{2\gamma})\pr= 1/F(f)$, then in a
	neghbourhood of $v^*= v(u^*)$ one has $v = \int F(f) df$.
	As $f$ tends to its finite (by (\ref{A5})) limit $f(v^*)$, this
	integral converges since, by (\ref{A6}), $F(f) \to 0$ in the same
	limit. Consequently, $|v^*| < \infty$, so that from $f' \to \infty$
	it follows $f'' \to \infty$ and hence $\RR_i \to \infty$ as
	$v\to v^*$, which proves Statement A1.

\medskip\noi
	Its immediate consequence is

\medskip\noi
	{\bf Statement A2.} \
	{\sl In static space-times with the metric (\ref{2.11}),
	where $ds_i^2$ are metrics of constant curvature spaces,
	a surface $u=u^*$ with $g_{tt}=0$ is a curvature singularity if its
	Hawking temperature $\TH$ is infinite. }

\medskip\noi
	Indeed, using e.g. formulae from the book \cite{Wald}, one finds
	for static metrics written in the form
\[
	ds^2 = -\e^{2\gamma(u)}dt^2 + \e^{2\alpha(u)}du^2 +
			\mbox{anything else}
\]
	the following expression for the Hawking temperature of a surface
	$u=u^*$ where $\e^{\gamma}=0$, assumed to be a horizon:
\beq
	\TH = \frac{1}{2\pi k_{\rm B}}\ \lim_{u\to u^*} \ \e^{\gamma-\alpha}
	           \left|\frac{d\gamma}{du}\right|              \label{THa}
\eeq
     where $k_{\rm B}$ is the Boltzmann constant. (The same expression can
	be obtained using other methods, such as the Euclidean continuation
	of the metric).

	We evidently have a special case of Statement 1, where, by the above
	notations, $\gamma = \beta_1$; more than that, it would be sufficient
	to have just $\e^{\gamma}<\infty$, while here, for an assumed horizon,
	a stronger condition $\e^{\gamma}=0$ is valid.

\medskip\noi
	{\bf Remark 5.} \ Statement A1 actually means, for spherical and other
	spatial symmetries, that any surface $u=u^*$ with $g_{tt}<\infty$ is a
	singularity if the gravitational force needed to keep a test particle
	at rest (proportional to $\e^{-\alpha}\dot\gamma$) is there
	infinite. Moreover, it can be easily shown that a surface $u=u^*$ with
	$g_{tt}=\infty$ is singular as well unless it is situated infinitely
	far from any external static observer (that is, unless the integral
	$\int \e^{\alpha}du$ diverges as $u\to u^*)$.  Other refinements of
	Statement 1 are possible but are here irrelevant.

\Acknow
{This work was supported in part by DFG grants
436 RUS 113/7, 436 RUS 113/236/O(R), by the Russian State Committee for
Science and Technology, and by the Russian Basic Research Foundation,
project N 95-02-05785-a.
K.B. acknowledges partial financial support from CAPES, Brazil,
and wishes to express his gratitude to colleagues from UFES, Brazil, for kind
hospitality.}

\end{document}